\documentclass[10pt,journal,compsoc]{IEEEtran}
\ifCLASSOPTIONcompsoc
  \usepackage[nocompress]{cite}
\else
  \usepackage{cite}
\fi
\usepackage[pdftex]{graphicx}

\usepackage{algorithm}
\usepackage{algpseudocode}
\usepackage{amsmath}

\usepackage{color}
\definecolor{ForestGreen}{RGB}{162,52,0}

\usepackage{xcolor}
\usepackage{amsmath}
\usepackage{CJK}
\usepackage{url}
\usepackage{array}

\usepackage{ragged2e}

\ifCLASSINFOpdf
\else
\fi

\begin{document}
%
\title{STEC-IoT: A Security Tactic by Virtualizing Edge Computing on IoT}

\author{Peiying Zhang, Chunxiao Jiang, ~\IEEEmembership{Senior Member,~IEEE}, Xue Pang and Yi Qian, ~\IEEEmembership{Fellow,~IEEE}
\IEEEcompsocitemizethanks{\IEEEcompsocthanksitem This work is partially supported by the National Natural Science Foundation of China under Grant 61922050, partially supported by the Major Scientific and Technological Projects of CNPC under Grant ZD2019-183-006, and partially supported by "the Fundamental Research Funds for the Central Universities" of China University of Petroleum (East China) under Grant 20CX05017A.\textit{(Corresponding author: Chunxiao Jiang.)}
}
\IEEEcompsocitemizethanks{\IEEEcompsocthanksitem Peiying Zhang is with the College of Computer Science and Technology, China University of Petroleum (East China), Qingdao 266580, China. \protect\\
E-mail: zhangpeiying@upc.edu.cn
}
\IEEEcompsocitemizethanks{\IEEEcompsocthanksitem Chunxiao Jiang is with the School of Information Science and Technology, Tsinghua University, Beijing 100084, China, with the Beijing National Research Center for Information Science and Technology, Beijing 100084, China. \protect\\
E-mail: jchx@tsinghua.edu.cn
}
\IEEEcompsocitemizethanks{\IEEEcompsocthanksitem Xue Pang is with the College of Computer Science and Technology, China University of Petroleum (East China), Qingdao 266580, China. \protect\\
E-mail: 1103746978@qq.com
}
\IEEEcompsocitemizethanks{\IEEEcompsocthanksitem Yi Qian is with Department of Electrical and Computer Engineering, University of Nebraska-Lincoln (UNL), United States. \protect\\
E-mail: yi.qian@unl.edu
}
\thanks{ }}

%
%

\markboth{IEEE Internet of Things Journal}%
{Shell \MakeLowercase{Chunxiao Jiang{et al.}}: STEC-IoT: A Security Tactic by Virtualizing Edge Computing on IoT}

\IEEEtitleabstractindextext{%
\begin{abstract}
\justifying\let\raggedright\justifying
To a large extent, the deployment of edge computing (EC) can reduce the burden of the explosive growth of the Internet of things. As a powerful hub between the Internet of things and cloud servers, edge devices make the transmission of cloud to things no longer complicated. However, edge nodes are faced with a series of problems, such as large number, a wide range of distribution, and complex environment, the security of edge computing should not be underestimated. Based on this, we propose a tactic to improve the safety of edge computing by virtualizing edge nodes. In detail, first of all, we propose a strategy of edge node partition, virtualize the edge nodes dealing with different types of things into various virtual networks, which are deployed between the edge nodes and the cloud server. Second, considering that different information transmission has different security requirement, we propose a security tactic based on security level measurement. Finally, through simulation experiments, we compare with the existing advanced algorithms which are committed to virtual network security, and prove that the model proposed in this paper has definite progressiveness in enhancing the security of edge computing.
\end{abstract}

}

\maketitle

\IEEEdisplaynontitleabstractindextext

\IEEEpeerreviewmaketitle

\section{Introduction}
\begin{figure}[!htp] 
\includegraphics[width=1.0\columnwidth]{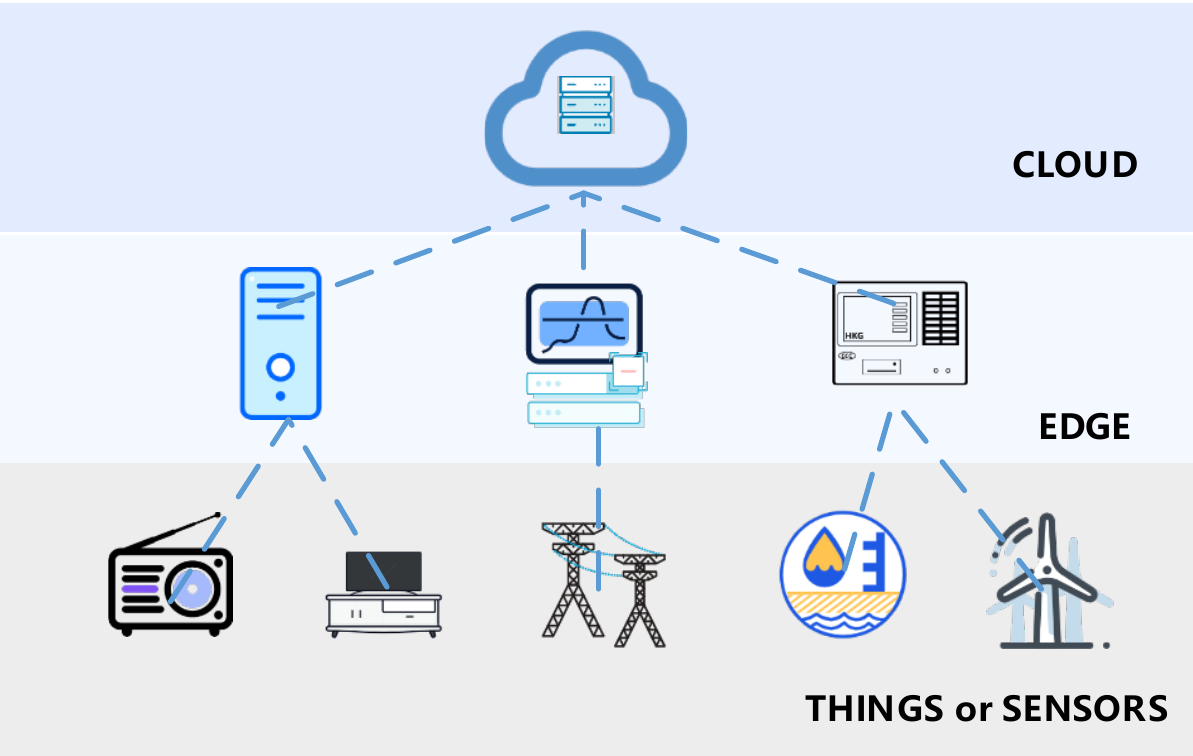}
\caption{Architecture of edge computing.}
\label{edge}
\end{figure}

The IoT is a giant-scale entirety composed of various networks \cite{DBLP:journals/iotj/WangZLBJ19}. Middleware in IoT can interact through different gateways\cite{DBLP:journals/jsac/JiangCLR13}. However, with the advent of the age of interconnection and the rapid popularization of wireless networks, the number of edge devices that make up the IoT has increased dramatically, and a large number of data information has been generated at the same time\cite{DBLP:journals/twc/JiangC0L13}. Although the traditional centralized public cloud service architecture has achieved some success, it still faces enormous challenges, because the concentration of resources will lead to longer delay and more periodic jitter on the network platform, while some real-time applications or delay-sensitive applications have higher requirements for low delay and jitter \cite{8812214}. As the intermediary between the cloud server and IoT, edge computing (EC) service for massive edge devices has played a permanent role in expanding and assisting cloud computing \cite{8489908}, the architecture of edge computing is shown in Fig.\ref{edge}.

The edge of the IoT architecture has many advantages, it can make the data generated by the data source be processed in the edge node, and solve the local business continuity problem. As the intermediate hub of cloud and things, EC mainly serves the downstream data from cloud servers and the upstream data from the IoT\cite{DBLP:journals/jsac/ZhuJKGL17}. A large number of requirements of different application scenarios are the core driving forces of the development of EC. In pursuance of meeting the low delay demands of delay-sensitive applications, EC can build solutions at the nearest location to avoid the delay caused by long-distance data transmission, in the massive data analysis business, EC can analyze and filter the data locally, avoiding the high cost caused by all the data transferred to the cloud.

The emergence of virtualization technologies is also to solve the problems brought by the rapid expansion of the IoT \cite{DBLP:journals/wc/WangWZG17}\cite{DBLP:journals/access/BizanisK16} \cite{8060514}, virtualization technologies can abstract all kinds of entity resources, and use them more flexibly according to their own needs. But how to combine EC and virtualization technologies for the work of the IoT needs more in-depth research, and the use of virtualization makes some potential security problems appear. In this paper, we combine EC with network virtualization (NV), discuss how to achieve a more efficient and secure cloud to things services. We summarize the main contributions of this paper below:

(1) According to the services provided by edge nodes, an edge node partition strategy is proposed, which uses slicing technology to classify the edge nodes first and establish a virtual network. Each virtual network is located between the edge node and the cloud server, and is responsible for different types of data requests and transaction processing.

(2) A security mechanism based on security level measurement is proposed, in which the virtual network and edge nodes are abstracted and established a virtual network mapping model, this model can distribute different kinds of services. It can effectively manage all types of businesses and avoid the confusion caused by the cross execution by different companies.

(3) A virtual network mapping model considering security requirements is introduced to deal with the question of data and service distribution from cloud to edge nodes. The model consists of node mapping stage and link mapping stage. During the node mapping phase, a multi-attribute comprehensive node mapping algorithm is proposed, it can map the virtual nodes to edge nodes satisfying different security requirements. During the link mapping phase, the links between virtual nodes are embedded to the links connecting edge nodes by particle swarm optimization algorithm.

(4) As far as we know, the model proposed in this article is unique among the existing studies. Through a series of simulation experiments, three existing algorithms considering the security of mapping process are compared. The experimental results verify the advanced nature of the proposed model in improving edge computing security, and solve the security problem of edge computing. The questions provide enlightenment.

The remainder of this paper is organized as follows. Section 2 reviews the existing methods for IoT and EC.Section 3 introduces the design of framework. In Section 4, we define the formula of evaluation index. In Section 5, we describe our proposed method in detail. The performance of our method and other methods is evaluated in Section 6. Section 7 concludes this paper.

\section{Related works}
In this chapter, we retrospect the forthcoming researches \cite{DBLP:journals/comcom/Zhang18} \cite{DBLP:journals/ppna/ZhangYLL19}\cite{DBLP:journals/jsac/WangLPHL20}, which can be divided into two parts. The first part introduces the existing work on IoT and EC\cite{DBLP:journals/jcp/XingLH13} \cite{XinboLiu} \cite{DBLP:journals/access/ZhangYL16} \cite{DBLP:journals/cn/LiuCXX15}, and the second part introduces the research on network virtualization \cite{DBLP:journals/iotj/ZhangYL18}.

\subsection{Researches on IoT and EC}

Researchers' research on edge computing extends to various fields, and the blockchain problem can also be solved by combining EC.
The authors of reference \cite{DBLP:journals/comsur/WangHLNYC20} discuss the relationship between mutually beneficial edge intelligence and intelligent edge, and introduce the relationship between them, such as application scenarios, technologies and challenges.
Literature \cite{DBLP:journals/iotj/XiongFWNWH19} and \cite{ZehuiXiong2020} respectively study the problems faced by public blockchain networks based on different consensus protocols. In literature \cite{ZehuiXiong2020} and \cite{DBLP:journals/iotj/XiongFWNWH19}, the computational problems in the process of public blockchain mining are studied by using the game theory of multi leader and multi follower.

The authors of literature \cite{DBLP:journals/network/XiongZLNWG20} put forward a method of using blockchain to support the data management of the IoT, and introduce a framework that can combine the IoT and blockchain.
The authors of reference \cite{DBLP:journals/network/WangHWZCC19} combine EC and deep learning to make edge systems more intelligent. They propose an "In-EDGE AI" framework, which can intelligently make devices and edge devices cooperate, and get better models while reducing communication burden,

At present, machine learning and network are closely combined to solve the optimization problems in the network environment \cite{DeepReinforcement2019}\cite{7792374}. In literature \cite{DeepReinforcement2019}, it is discussed that with the development of 5G network, the network becomes more and more intensive, and different services are difficult to be guaranteed. The authors of this paper introduce a method to solve local decision-making through deep reinforcement learning (DRL). The idea is to let participants in the network learn and build the network, and then make the best decision. Committed to supporting the security and QoS of IOV, a vehicle Internet use blockchain is proposed in \cite{DBLP:journals/tvt/KangXNYKZ19}. Document \cite{DBLP:journals/wc/WangWZG17} introduce a new method for the future network, which can solve the complex interface and backhaul problems between densely distributed cells.

As a part of the IoT, the mobile social Internet of things (SIoT) has also received the attention of researchers. In the literature \cite{8241407}, a data unloading method is proposed, which can realize data transactions of different users on the mobile social platform. In the literature \cite{7397054}, the authors use the software definition network (SDN) with virtualization structure to classify and virtualize the facilities in the social Internet of things.

\subsection{Researches on network virtualization}
VNE mapping in the substrate domain has been well studied, but the demand for virtual networks is growing rapidly, which requires a wide range of virtual networks across multiple domains \cite{8485467}\cite{DBLP:journals/access/ZhangYQL18}\cite{9096506}\cite{DBLP:journals/cn/BesiktasGUL17}\cite{DBLP:journals/jnsm/ZhangLNGLW20}. At present, the research on this problem pays less attention to the joint relationship between intra domain links and inter domain links, mainly focusing on decomposing a VN into subvns of each domain.

Attacking and tampering with the Internet is a serious crime, this kind of behavior will violate security attributes such as confidentiality and integrity. Therefore, it is very important for virtualization architecture to provide protection against these threats and other threats that may endanger its security. In order to be able to use virtualization in a real environment, effective resource allocation and security regulations must be considered at the same time. To solve this problem, the authors of literature \cite{DBLP:conf/cnsm/BaysOBBG12} introduce a new way that can optimize substrate resources and security performance. In order to ensure the transmission delay and security, the authors of reference \cite{9094032} propose a cooperation strategy between nodes, which can reduce the average delay and hit rate of downlink.

The authors of literature \cite{shuiqinggong} introduce trust relationship and trust degree, build a mathematical model of security virtual network mapping, and propose a trust aware security virtual network mapping algorithm. In the mapping process, the importance of multiple attributes of nodes is sorted by using the approach ideal ordering method, so that the importance of local and global nodes is well considered. However, the optimization of link mapping is not considered in this algorithm. The authors of reference \cite{9067004} model the three-layer heterogeneous satellite network as a network model, using time expansion graph, which greatly reduced the search space and improved the efficiency of the algorithm.

In \cite{Penglimin2015}, the authors propose a minimum cost embedding algorithm. Firstly, according to the constraints of virtual network requests, the candidate substrate node set is constructed,and then, the available substrate path set of each virtual link is calculated.
\section{Architecture design}

\subsection{Edge node partition strategy}
\begin{figure}[!htp] 
\centering
\includegraphics[width=1.0\columnwidth]{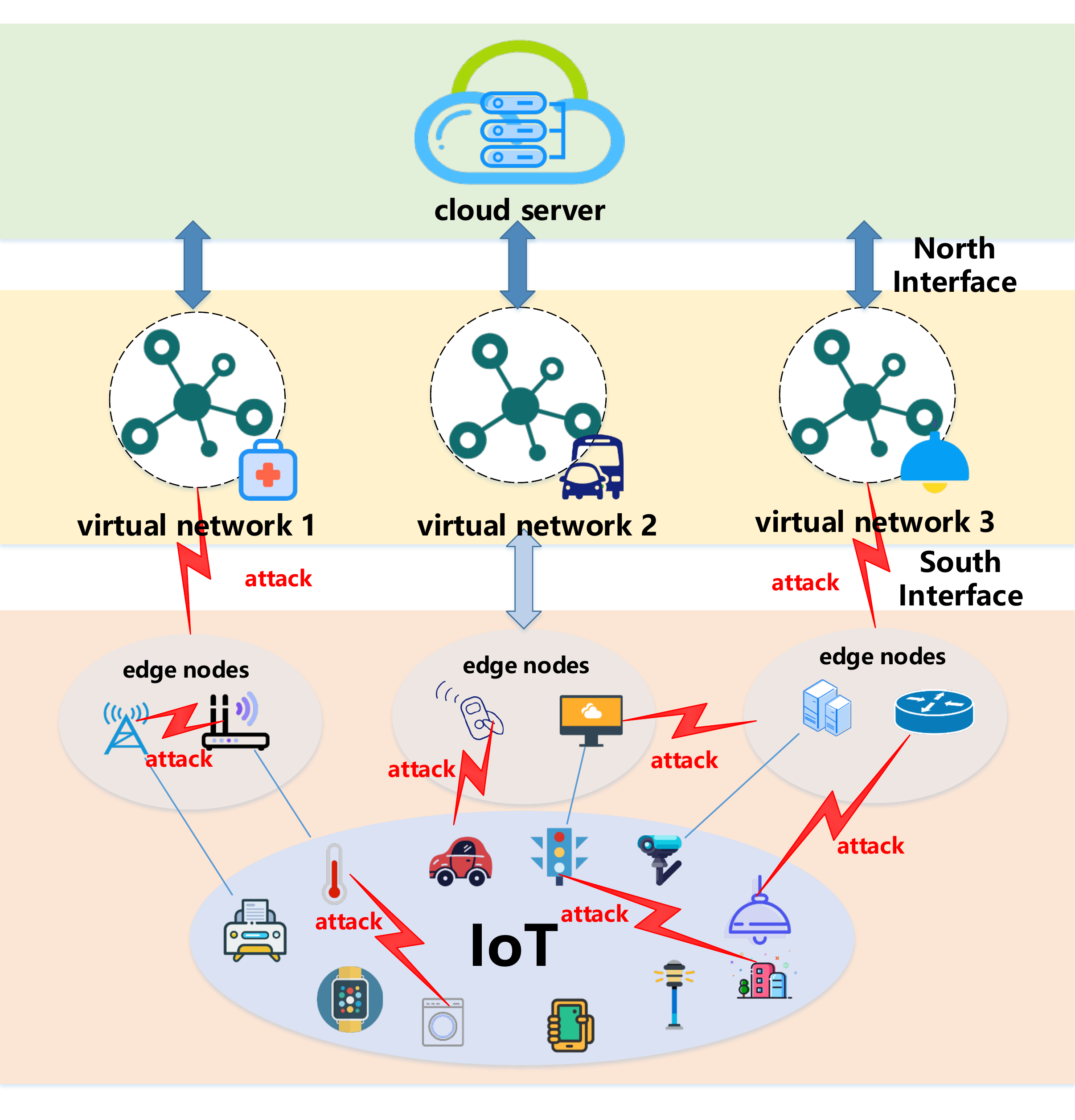}
\caption{The virtualized edge computing scenario.}
\label{iot}
\end{figure}

In this chapter, we propose a strategy to virtualize edge nodes: edge node partition strategy. It can classify the edge nodes according to specific rules, and then abstract all kinds of edge nodes into different virtual networks. The virtualized edge computing model is shown in Fig.\ref{iot}.

Take Fig.\ref{iot} as an example to introduce how the strategy works. First of all, each edge node is responsible for the closest physical facilities on the Internet of things. We classify these edge nodes according to service types. As shown in the figure, we divide the edge nodes serving medical facilities into the first category, the edge nodes serving transportation facilities into the second category, and the edge nodes serving household facilities into the third category. There are more kinds of practical applications. Here we only use three examples to illustrate the situation. In this way, edge nodes scattered in different locations are divided together due to similar characteristics.

The virtual network can meet different service requests at the same time. Here, we virtualize the edge nodes which are divided into different classes, that is, the same kind of edge nodes are abstracted into a virtual network, and each virtual network is responsible for distributing and filtering different data and service requests. The virtual network is located among the cloud server and the edge nodes. It has two interfaces. The north interface is linked to the cloud server, which is used to receive and process the downlink data from the cloud server. The south interface is connected to edge nodes, which is used to process the uplink data from edge nodes.

\subsection{Security mechanism based on security level measurement}
In the process of data distribution and receiving, all kinds of facilities or users in the network are vulnerable to attack, which is also the main problem considered in this paper. Security risks can be divided into three types: (1) mutual attacks between virtual network and edge nodes; (2) mutual attacks between edge nodes; (3) mutual attacks between virtual nodes. Considering the security of data transmission, we propose a security mechanism based on the security level measurement, which abstracts the virtual network and edge network into a network composed a great many of nodes, and models the data transmission process as a virtual network embedding (VNE) model,  Fig.\ref{network_model} shows the model:
\begin{figure}[!htp] 
\includegraphics[width=1.0\columnwidth]{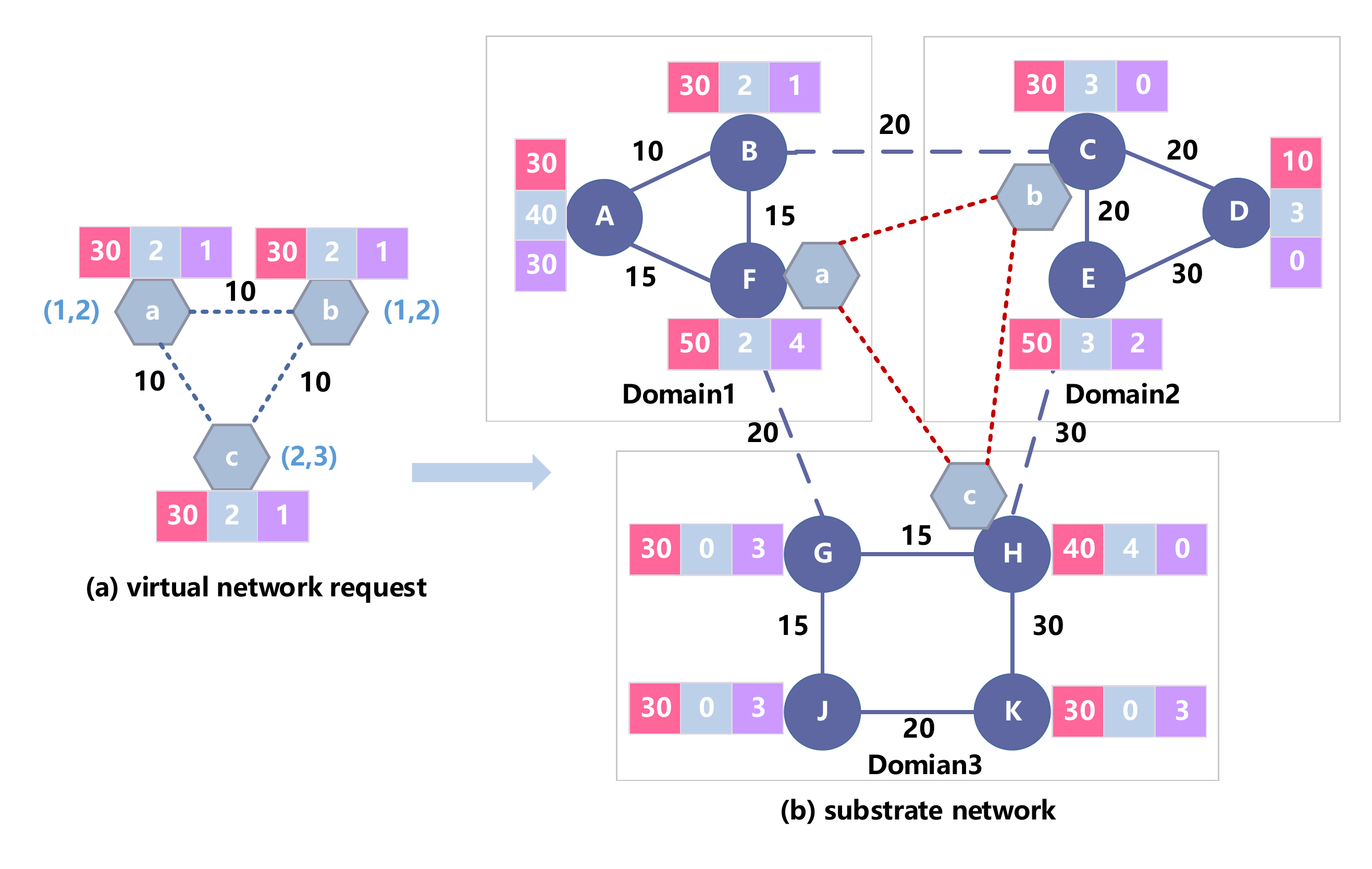}
\caption{The model of virtual network and substrate network.}
\label{network_model}
\end{figure}

\textbf{Virtual network model}
The virtual network abstracted from edge nodes can be modeled as an undirected weighted graph $G_{V}= \left ( N_{V},L_{V} \right )$, where $N_{V}= \left \{ n_{v}^{i} \mid i= 1,2,3 \cdots \right \}$ represents the set of virtual nodes, $L_{V}= \left \{ l_{v}^{j,k} \mid n_{v}^{j}\in N_{V},n_{v}^{k}\in N_{V}\right \}$ represents the set of virtual links, and the attributes of virtual node $n_{v}^{i}\in N_{V}$  include cpu resources demand $cpu\left ( n_{v}^{i} \right )$, the set of candidate domains $cd\left ( n_{v}^{i} \right )$, security demand level $vsd\left ( n_{v}^{i} \right )$, security level $vsl\left ( n_{v}^{i} \right )$, and the attributes of virtual link $l_{v}^{jk}\in L_{V}$ include bandwidth resources demand $bw\left ( l_{v}^{jk} \right )$.

A graph of virtual network request is shown in Fig.\ref{network_model}(a). The hexagon represents the virtual node, the dotted line connecting the nodes represents the virtual link, the attributes of the virtual node are represented by three squares, the number in the first square represents the CPU resources demand of the virtual node, the number in the second square represents the security demand level of the virtual node, the security level of the virtual node is represented by a number in the third square. The bandwidth demand of the virtual link is indicated next to the link.

\textbf{Substrate network model}
The substrate network composed of edge nodes can also be modeled as an undirected weighted graph $G_{S}= \left ( N_{S},L_{S} \right )$, where $N_{S}$ represents the set of substrate nodes, $L_{S}$ represents the set of substrate links, we assume that the entire substrate $G_{S}$ is composed of $N$ domains managed by $N$ InPs, which are interconnected by multiple inter-domain links. The $k$-th substrate domain can be defined as $G_{S,K}=(N_{S,K}, L_{S,K})$, where $N_{S,K}=\left \{ n_{s,k}^{i} \mid i= 1,2,3 \cdots \right \}$ and $L_{S,K}=\left \{ l_{s,k}^{jt} \mid n_{s,k}^{j}\in N_{S,K},n_{s,k}^{t}\in N_{S,K}\right \}$ represent the set of substrate nodes and substrate links which are managed by $k$-th InP. In addition, we define the set of inter-domain substrate links as $E_{S,I}$. Thus, $G_{S}$ can be formulated as:

\begin{equation}
\begin{aligned}
G_{S}=G_{S,1} \cup G_{S,2} \cup ... \cup G_{S,N} \cup E_{S,I}, \\
N_{S}=N_{S,1} \cup N_{S,2} \cup N_{S,3} ... \cup N_{S,N}, \\
L_{S}=L_{S,1} \cup L_{S,2} \cup ... \cup L_{S,N} \cup E_{S,I}.
\end{aligned}
\end{equation}

The attributes of substrate node $n_{s,k}^{i}\in N_{S,K}$ include available cpu resources $cpu\left ( n_{s,k}^{i} \right )$, security demand level $ssd\left ( n_{s,k}^{i} \right )$, security level $ssl\left ( n_{s,k}^{i} \right )$, and the attributes of substrate link $l_{s,k}^{jt}\in L_{S,K}$ include available bandwidth resources $bw\left ( l_{s,k}^{jt} \right )$.

As shown in Fig.\ref{network_model} (b), a substrate network consists three domains, each domain is represented by an ellipse, the substrate node is represented by a circle, their attribute values are displayed in three squares, the number in the first square represents the available CPU  resources of the substrate node, the number in the second square represents the security demand level of the substrate node, and the security level of the substrate node is represented by the number in the third square. The solid line between each substrate node represents the intra-link, the inter-link is represented by dotted line, and the available bandwidth resources of each substrate link is marked next to it.

\textbf{Virtual network embedding model of EC}
We express the mapping process by $G_{V}= \left ( N_{V},L_{V} \right )\rightarrow G_{S}= \left ( N_{S},L_{S} \right )$. According to the constraints, virtual nodes a, b, c are mapped to physical nodes F, C, H respectively.

\section{mathematical model}
We model the process as a mixed-integer linear program (MILP) model. With satisfying the constraints of node and link resources, we solve the problem by minimizing the embedding consumption of substrate network as the objective function. The detailed explanation is below:
\subsection{evaluation metrics}
We use the percentage of successfully mapped virtual network requests and all virtual network requests within a certain time interval to represent the acceptance rate. It can be expressed as:
\begin{equation}
\label{eq1:acceptancerate}
AcpRate\left ( G_{v},t_{a},t_{b} \right )= \frac{\sum_{t_{a}}^{t_{b}}accept\left ( VNR \right )}{\sum_{t_{a}}^{t_{b}}all\left ( VNR \right )},
\end{equation}
wherein, $\sum_{t_{a}}^{t_{b}}accept\left ( VNR \right )$ represents the virtual network requests mapped successfully between $t_{a}$ and $t_{b}$, and $\sum_{t_{a}}^{t_{b}}all\left ( VNR \right )$ represents all virtual network requests between $t_{a}$ and $t_{b}$.

The embedding revenue of virtual network request form $t_{a}$ to $t_{b}$ can be expressed as $Rev\left ( G_{v},t_{a},t_{b} \right )$, and its calculation formula is as follows:
\begin{equation}
\label{eq2:revenue}
Rev\left ( G_{v},t_{a},t_{b} \right )=\alpha\sum_{n_{v}^{i}\in N_{V}} cpu\left ( n_{v}^{i} \right )+\beta\sum_{l_{v}^{jk}\in L_{V}} bw\left ( l_{v}^{jk} \right ),
\end{equation}
\begin{equation}
\label{eq2:abweight}
\alpha +\beta = 1,
\end{equation}
wherein, $\sum_{n_{v}^{i}\in N_{V}} cpu\left ( n_{v}^{i} \right )$ represents the sum of CPU resources demand of the virtual request, $\sum_{l_{v}^{jk}\in L_{V}} bw\left ( l_{v}^{jk} \right )$ represents the sum bandwidth resources demand of virtual links. $\alpha$ and $\beta$ are parameters with weight, they distinguish the extent of nodes and links.

In order to facilitate comparison, we present the unit embedding revenue of the virtual network request between $t_{a}$ and $t_{b}$ as follows:
\begin{equation}
\label{eq3:avg_revenue}
Rev_{avg}\left ( G_{v},t_{a},t_{b} \right )= \frac{Rev\left ( G_{v},t_{a},t_{b} \right )}{t_{a}-t_{b}}.
\end{equation}

The embedding consumption of virtual network request form $t_{a}$ to $t_{b}$ can be expressed as $Cons\left ( G_{V},t_{a},t_{b} \right )$, and the calculation formula is as follows:

\begin{equation}
\label{eq4:consumption}
\begin{aligned}
Cons\left ( G_{V},t_{a},t_{b} \right )=\omega _{1}\left (n_{v}^{i},n_{s}^{m} \right )\sum_{n_{v}^{i}\in N_{V}}cpu\left (n_{v}^{i} \right )\\
+\omega _{2}\left ( l_{v}^{jk},l_{v}^{pq} \right )\sum_{l_{v}^{jk}\in L_{V}}bw\left ( l_{v}^{jk} \right ),
\end{aligned}
\end{equation}
wherein, $\sum_{n_{v}^{i}\in N_{V}} cpu\left ( n_{v}^{i} \right )$ represents sum of CPU resources demand of the virtual request, and $\sum_{l_{v}^{jk}\in L_{V}} bw\left ( l_{v}^{jk} \right )$ represents the sum bandwidth resources demand of virtual links. There are two values of $\omega _{1}\left (n_{v}^{i},n_{s}^{m} \right )$. If there is a mapping relationship between $n_{v}^{i}$ and $n_{s}^{m}$, take 1. If there is no mapping relationship between, take 0. So is $\omega _{2}\left ( l_{v}^{jk},l_{v}^{pq} \right )$.

As a matter of convenience, we present the unit mapping consumption of the virtual network request as follows:
\begin{equation}
\label{eq5:avg_consumption}
Cons_{avg}\left ( G_{V},t_{a},t_{b} \right )=\frac{Cons\left ( G_{V},t_{a},t_{b} \right )}{t_{a}-t_{b}}.
\end{equation}

The revenue-consumption ratio of virtual network request is expressed by unit mapping benefit and unit mapping cost, which can better reflect the mapping effect of virtual network request.
\begin{equation}
\label{eq6:revenue_consumption}
\frac{R}{C}=\frac{Rev_{avg}\left ( G_{V},t_{a},t_{b} \right )}{Cons_{avg}\left ( G_{V},t_{a},t_{b} \right )}.
\end{equation}

\subsection{objective function}
We take minimizing the consumption of substrate network embedding as the objective function, which is expressed as follows:
\begin{equation}
\label{objective function}
\begin{aligned}
\min [ \omega _{1}\left (n_{v}^{i},n_{s,k}^{m} \right )\sum_{n_{v}^{i}\in N_{V}}cpu\left (n_{v}^{i} \right )\\
+\omega _{2}\left ( l_{v}^{jk},l_{s,k}^{pq} \right )\sum_{l_{v}^{jk}\in L_{V}}bw\left ( l_{v}^{jk} \right )].
\end{aligned}
\end{equation}

\subsection{constraint condition}
In the mapping process, the mapping can be completed only when the constraints of the mapping are met. The mapping conditions required are described below:

The node constraints are shown below:
\begin{equation}
\begin{split}
\label{constraint1}
\omega _{1}\left (n_{v}^{i},n_{s,k}^{m} \right )cpu\left (n_{v}^{i} \right )\leq cpu\left (n_{s,k}^{m} \right ),\\
n_{v}^{i}\in N_{V},n_{s,k}^{m}\in N_{S},
\end{split}
\end{equation}

\begin{equation}
\label{constraint2}
\sum\omega _{1}\left (n_{v}^{i},n_{s,k}^{m} \right )=1,n_{v}^{i}\in N_{V},n_{s,k}^{m}\in N_{S},
\end{equation}

\begin{equation}
\label{constraint3}
\omega _{1}\left (n_{v}^{i},n_{s,k}^{m} \right )=\left\{\begin{matrix} 1 ,& n_{s,k}^{m}\in cn\left ( n_{i}^{i} \right )
\\ 0 ,& n_{s,k}^{m}\notin  cn\left ( n_{i}^{i} \right )
 \end{matrix}\right..
\end{equation}

Equation (\ref{constraint1}) indicates that the available CPU resources of the mapped substrate nodes must be greater than or equal to the CPU resources demand of virtual nodes. Equation (\ref{constraint2}) indicates that a physical node can only host one virtual node. Equation (\ref{constraint3}) indicates that the value of $\omega _{1}\left (n_{v}^{i},n_{s,k}^{m} \right )$ can only be 1 and 0, when the substrate node belongs to the candidate domain set, the value of $\omega _{1}\left (n_{v}^{i},n_{s,k}^{m} \right )$ is 1, when the substrate node does not belong to the set, the value of $\omega _{1}\left (n_{v}^{i},n_{s,k}^{m} \right )$ is 0.

The link constraints are shown below:

\begin{equation}
\label{constraint4}
\sum bw\left ( l_{v}^{jk} \right )\leq bw\left ( l_{s,k}^{pq} \right ),l_{v}^{jk}\in L_{V},l_{s,k}^{pq}\in L_{S,K},
\end{equation}

\begin{equation}
\label{constraint5}
\omega _{1}\left ( n_{v}^{i},n_{s,k}^{p} \right )-\omega _{1}\left ( n_{v}^{j},n_{s,k}^{p} \right )\leq 1,
\end{equation}

\begin{equation}
\label{constraint6}
\omega _{2}\left ( l_{v}^{ij},l_{s,k}^{pq} \right )\in \left \{ 0,1 \right \}.
\end{equation}

Equation (\ref{constraint4}) illustrates that the total bandwidth of the virtual links that be mapped must be smaller than or equal to the bandwidth of the substrate link. Equation (\ref{constraint5}) claims that the virtual link cannot be mapped to a ring-shaped substrate link, that is, the two ends of the virtual link cannot be mapped to the same substrate node. Equation (\ref{constraint6}) indicates that a virtual link can only be mapped to one substrate link.

The security constraints are shown below:

\begin{equation}
\label{constraint7}
ssd\left ( n_{s,k}^{j} \right )\leq vsl\left ( n_{v}^{i} \right ),
\end{equation}

\begin{equation}
\label{constraint8}
vsd\left ( n_{v}^{i} \right )\leq ssl\left ( n_{s,k}^{j} \right ).
\end{equation}

Equation (\ref{constraint7}) follows the rules that the sdl of the substrate node less than or equal to the sl of virtual node it hosted, and equation (\ref{constraint8}) follows the rules that the sdl of virtual node must less than or equal to the sl of substrate node it mapped.

\section{algorithm design}

\subsection{node mapping stage}
In the node mapping phase, due to the high security requirement of security virtual network mapping, we define a virtual node priority formula to calculate the priority of each virtual node:
\begin{equation}
\label{nodemapping1}
Priority\left ( n_{v}^{i} \right )= vsd\left ( n_{v}^{i} \right )*cpu\left ( n_{v}^{i} \right ).
\end{equation}

Formula (\ref{nodemapping1}) shows that when the security demand level of virtual nodes is high or the resources demand of CPU is high, in order to ensure the security and resources demands of the nodes, the virtual nodes with higher priority function value are selected for mapping.

All the virtual node are sorted according to the priority of them. First, according to the defined constraints, we select the candidate substrate nodes. The candidate physical nodes set can be expressed as follows:
\begin{equation}
\label{nodemapping2}
\begin{aligned}
CN\left ( n_{v}^{i} \right )= \{ n_{s,k}^{m}\mid k\in cd\left ( n_{v}^{i} \right ),\\
cpu\left ( n_{s,k}^{m} \right ) > cpu\left ( n_{v}^{i} \right ),\\
ssl\left ( n_{s,k}^{m} \right )> vsd\left ( n_{v}^{i} \right ),\\
vsl\left ( n_{s,k}^{m} \right )> ssd\left ( n_{v}^{i} \right )\}.
\end{aligned}
\end{equation}

This constraint indicates that the candidate substrate node must belong to the candidate domain set, the CPU resources of the substrate node must be greater than that of the virtual node, the secure demand level of the substrate node should be less than secure level of it, and the secure level of the substrate node should be greater than secure demand level of the virtual node. Only when the above conditions are met can the candidate substrate node be considered.
In order to select the candidate substrate nodes more equitably, we define a substrate node priority function as follows:
\begin{equation}
\label{nodemapping3}
\begin{aligned}
Priority\left ( n_{s,k}^{m} \right )= \gamma \left [ ssl\left ( n_{s,k}^{m} \right )-vsd\left ( n_{v}^{i} \right ) \right ]\\
+\delta \left [ cpu\left ( n_{s,k}^{m} \right )-cpu\left ( n_{v}^{i} \right ) \right ]+\theta hop\left ( n_{s,k}^{m} \right ).
\end{aligned}
\end{equation}

In this formula, the first part represents the difference between the security level of the substrate node and the security demand level of the virtual node, the second part represents the difference between the available CPU resources of the substrate node and the CPU resources demand of the virtual node, and the third part represents the distance between the substrate node and the boundary node. If the substrate node is the boundary node, the value is 0. The coefficient $\gamma$, $\delta$, $\theta$ represent the weight of each part, and their values are 0.5, 0.3 and 0.2 respectively. The reason for setting the weight is that the priority should be given to the security attributes. When the security attributes are the same, the CPU resources should be considered. This can effectively avoid fragmentation. When the above two conditions are the same, the physical nodes close to the boundary nodes should be selected for mapping.

The pseudo code of the node mapping algorithm is shown in \textbf{Algorithm \ref{nodemapping}}.
\begin{algorithm}[h]
  \caption{Node Mapping Algorithm}
  \label{nodemapping}
  \begin{algorithmic}[1]
    \For {each virtual node}
    \State calculate the $Priority\left ( n_{v}^{i} \right )$;
    \State prioritize all virtual nodes;
    \State select $Candi\_node\left ( n_{v}^{i} \right )$;
    \For {each virtual node}
    \State calculate the $Priority\left ( n_{s,k}^{m} \right )$;
    \EndFor
    \EndFor
    \State map the node according to $Priority\left ( n_{v}^{i} \right )$ and $Priority\left ( n_{s,k}^{m} \right )$;
    \State \Return node mapping result;
  \end{algorithmic}
\end{algorithm}

\subsection{Virtual network embedding using PSO algorithm}
Particle swarm optimization (PSO) is a bionic optimization algorithm based on multiple agents. The basic core of it is that the agents in the group cooperate with each other to better adapt to the environment. The inertia weight model of PSO is as follows:

\begin{equation}
\begin{split}
\label{pso1}
V_{i+1}=\omega V_{i}\left ( t \right )\oplus r1*c1\left (X _{pb} \ominus X _{i} \right )\\
\oplus r2*c2\left (X _{gb}\ominus X _{i} \right ),
\end{split}
\end{equation}

\begin{equation}
\label{pso2}
X_{i+1}=X_{i} \otimes V_{i+1}.
\end{equation}

We use vector to represent each particle, that is, a specific mapping scheme, and component to represent the physical nodes mapped by each virtual node.

Next, we explain the parameters in the formula:

Particle position $X_{i}$: $X_{i}=\left \{ x_{i}^{1},x_{i}^{2},x_{i}^{3},\cdots ,x_{i}^{n} \right \}$ represents the $i$th mapping scheme, $n$ represents the number of virtual nodes in the request scheme, and $x{i}^{k}$ represents the number of physical node mapped by the $k$th virtual node.

Particle velocity $V_{i}$: $V_{i}=\left \{ v_{i}^{1},v_{i}^{2},v_{i}^{3},\cdots ,v_{i}^{n} \right \}$ indicates whether to change the $i$th mapping scheme, where the value of $v_{i}^{k}$ is 0 or 1. When its value is 0, it indicates that the virtual node needs to select a new physical node for mapping. When its value is 1, it indicates that the mapping scheme is unchanged.

Subtraction $\ominus$: Used to calculate whether the two embedding schemes are the same, when the two items are the same, take 1; when the two items are different, take 0.

Addition $\oplus$: After adding by bits, carry out rounding operation.

Multiplication $\otimes$: Decides whether to re-select the candidate node.

\textbf{Algorithm \ref{psomapping}} describes the process of mapping with PSO.
\begin{algorithm}[h]
  \caption{VNE algorithm using PSO}
  \label{psomapping}
  \begin{algorithmic}[1]
    \For {each virtual node}
      \State initialize $X_{i}$ and  $V_{i}$;
    \EndFor
    \For {$i<N$}
      \For {$j<M$}
      \State update particle position $X_{i+1}$;
      \State update particle velocity $V_{i+1}$;
      \If {$f(X_{pb})>f(X_{i+1})$}
        \State $X_{pb}=X_{i+1}$
      \EndIf
      \If {$f(X_{gb})>f(X_{pb})$}
        \State $X_{gb}=X_{pb}$
     \EndIf
      \EndFor
    \EndFor
  \State \Return mapping result;
  \end{algorithmic}
\end{algorithm}

\section{Simulation Experiment and Analysis}
The environment used in this paper is 4GB memory, 64 bit windows 10 operating system. We write the simulation program in Java language and run it in eclipse. We randomly generate the topologies for simulation experiments on eclipse. Finally, the comparison chart of experimental results is drawn with origin. See Table 1 for simulation experiment parameters.

\begin{table}[htpb]
\caption{The Settings of Parameters}
\label{tbl_parameters_setting}
\begin{tabular}{lm{3cm}cm{2cm}}
\hline
Parameter Items                      & The Range    \\ \hline
the amount of substrate domains      & 4            \\
the amount of physical nodes        & 120          \\
substrate node cpu resources             & U[0,50]   \\
intra-domain link connection rate    & 0.6      \\
substrate link bandwidth resources             & U[1000,3000] \\
ssl of the substrate node              & U[0,4]\\
ssd of the substrate node              & U[0,4]\\ \hline
virtual nodes number       & U[2,10]      \\
virtual node  cpu resources        & U[50,100]      \\
virtual link bandwidth resources          & U[1,10]      \\
vsd of the virtual node              & U[0,4]\\
vsd of the virtual node              & U[0,4]\\
parameter $\alpha$               & 0.5          \\
parameter $\beta$               & 0.5          \\ \hline
the number of particles in PSO & 10\\
iterations in PSO & 50\\
parameter $\omega$ & U[0.1,0.9]\\
parameter $r1$ and $r2$ & U[0,1]\\
parameter $c1$ and $c2$ & 1.5\\
\hline
\end{tabular}
\end{table}

\subsection{Analysis of experimental results}
In this section, the STEC-IoT algorithm proposed in this paper and three other algorithms are compared, the TA-SVNE \cite{gongshuiqing} algorithm is a VNE algorithm which uses the trust perception of security. The greedy strategy is introduced in the G-SVNE \cite{gongshuiqing} algorithm. The MCS-VNE \cite{penglimin} is a new MC-VNE adding a secure attribute, we compare the three algorithms from four aspects: the acceptance rate of virtual network request, the revenue of substrate network, the cost of substrate network and the revenue/cost ratio of substrate network.

\begin{table}[htpb]
\caption{The compared four methods.}
\label{tbl_parameters_setting}
\begin{tabular} {|p{1.5cm}|p{6.5cm}|} \hline
Notation & The idea of algorithm \\
\hline
STEC-IoT& \multicolumn{1}{|m{6.5cm}|} {In this algorithm, the candidate physical nodes are selected by considering multiple attributes, and then the nodes are mapped according to the resource metrics of nodes. In the process of link mapping, PSO algorithm is used to optimize the mapping process, and the final mapping results are obtained.}\\\hline
TA-SVNE& \multicolumn{1}{|m{6.5cm}|} {In the mapping process, the algorithm considers the local and global importance of the nodes, and uses the TOPSIS to map the nodes according to their importance, and then K-shortest path algorithm is used for link mapping.}\\\hline
G-SVNE& \multicolumn{1}{|m{6.5cm}|} {In this algorithm, the greedy strategy is used to get the optimal solution of each step, and then the solution of each step is combined to get the final mapping result.}\\\hline
MCS-VNE& \multicolumn{1}{|m{6.5cm}|} {This algorithm is obtained by MC-VNE algorithm adding security attributes. The specific process is explained in section 2.}\\\hline
\end{tabular}
\end{table}

\textbf{Experiment 1:} The comparison of virtual network request acceptance rate.

\begin{figure}[!htp] 
\includegraphics[width=1.0\columnwidth]{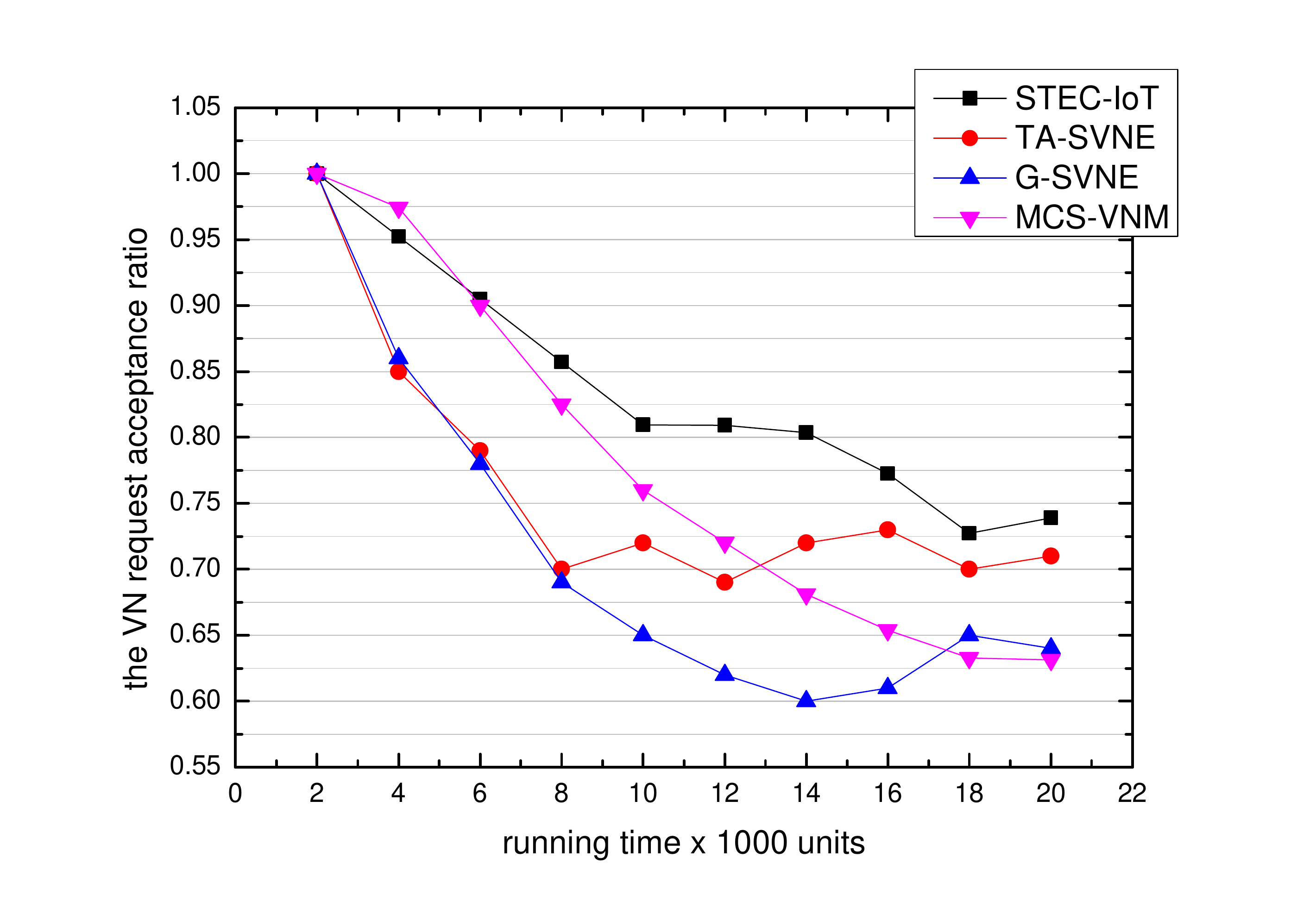}
\caption{The virtual network request acceptance rate.}
\label{ex1}
\end{figure}

Fig.\ref{ex1} shows the results of Experiment 1. The figure shows the difference of request acceptance rate among the four algorithms. We can see in the picture, in general, four algorithms of data over time, have a downward trend, this is because as the growth of the time, the available resources of substrate network are less and less, so there are a lot of virtual network request cannot meet the demand of resources, lead to the failure mapping. In the prophase, the acceptance rate of STEC-IoT algorithm introduced by us is better than that of the other three methods, in the later, the acceptance rate of STEC-IoT algorithm decreased, but still remain above 0.6. The data of G-SVNE and TA-SVNE is not stable enough, even as low as 0.6. This is because the greedy algorithm only considers the optimal choice without comprehensive consideration of many aspects, resulting in the lowest acceptance rate.

\textbf{Experiment 2:} The comparison of substrate network revenue.

\begin{figure}[!htp] 
\includegraphics[width=1.0\columnwidth]{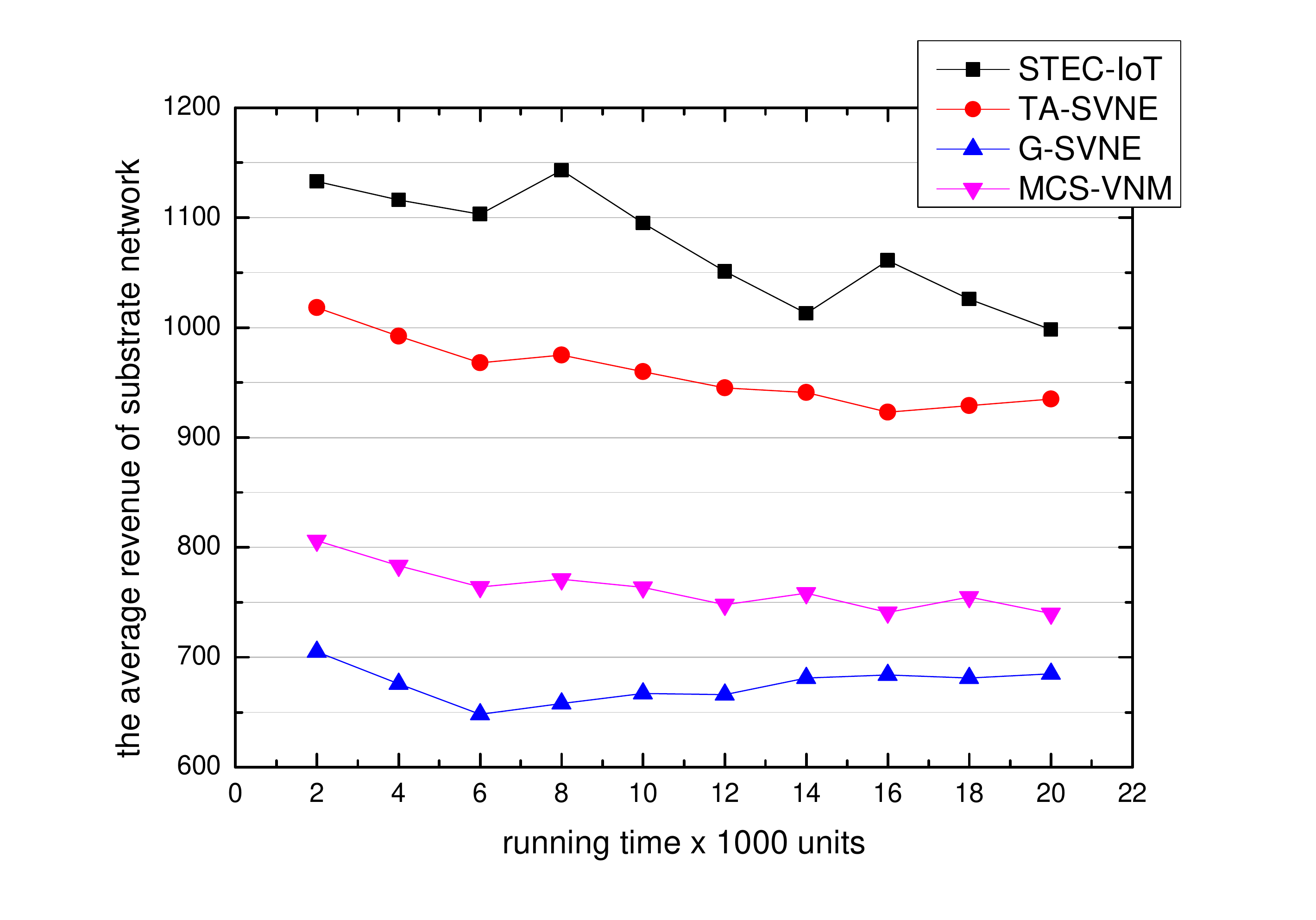}
\caption{The revenue of substrate network.}
\label{ex2}
\end{figure}

Fig.\ref{ex2} clearly shows the revenue changes of the four algorithms. As shown in the figure, the mapping revenue of STEC-IoT algorithm and TA-SVNE algorithm have a higher value than that of G-SVNE algorithm, while the mapping revenue of MCS-SVNE algorithm has a higher value than that of G-SVNE, which is attributed to appropriate node and link embedding method. As the time becomes longer and longer, the values of the four curves grow slowly. There are two reasons for the slow growth, the first one is that the targets selected in the adjacent time are similar by using the same equation, they have detailed resources, the second one is that as more and more nodes and links are mapped, the remaining available resources are consumed, and the physical resources that can meet the conditions are less and less, the above two reasons hinder the rapid growth of the curve. Because STEC-IoT algorithm has the highest acceptance rate of virtual network requests, the number of virtual networks mapped successfully in the same time is the largest, and the average income of physical network is the highest.

\textbf{Experiment 3:} The comparison of substrate network cost.

\begin{figure}[!htp] 
\includegraphics[width=1.0\columnwidth]{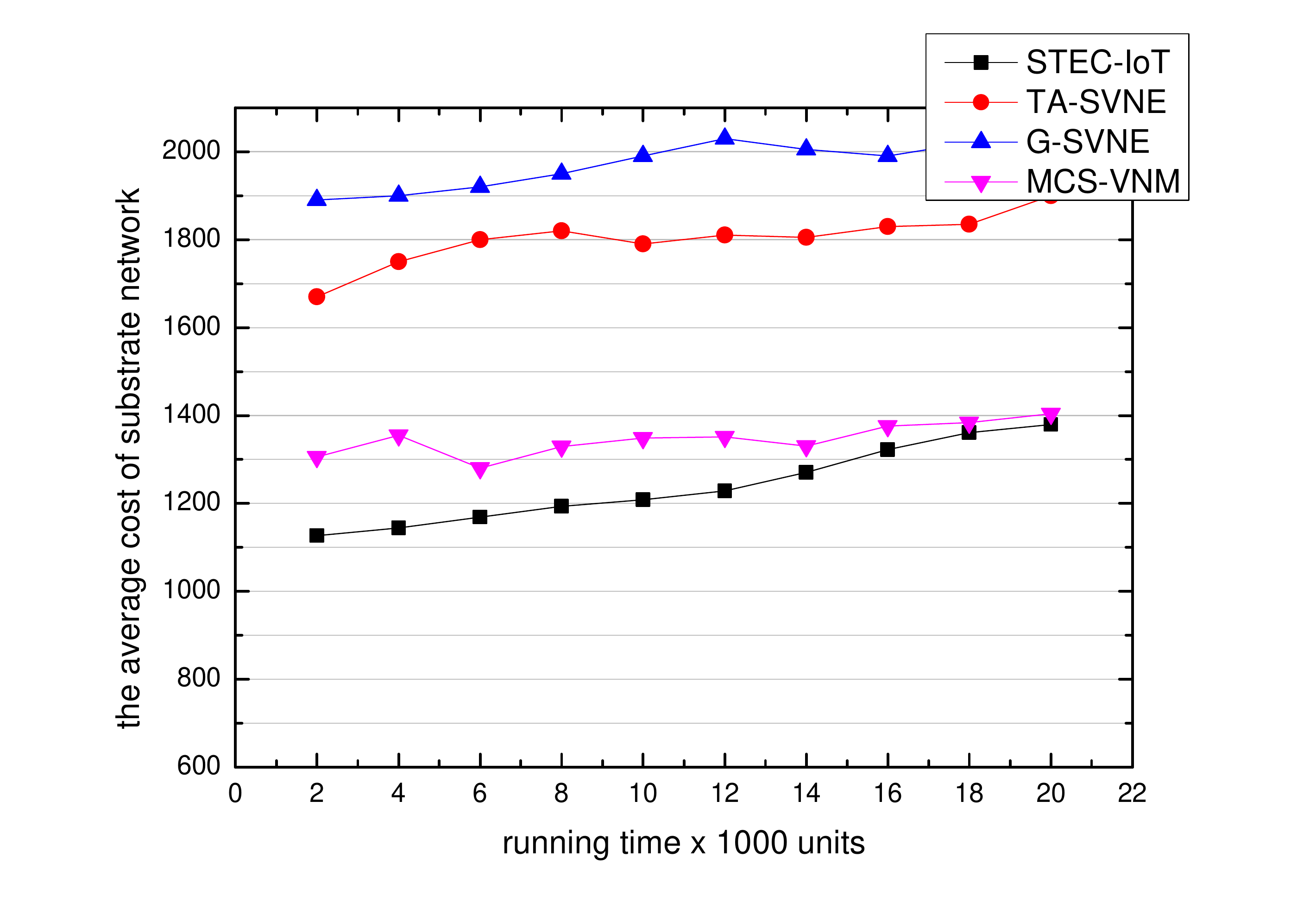}
\caption{The cost of substrate network.}
\label{ex3}
\end{figure}

The four curves in Fig.\ref{ex3} represent the cost values of the four algorithms, and each curve is rising steadily. This is because in the early stage of simulation experiment, the number of virtual network requests arriving is small, so the mapping cost is low. With the increase of simulation time, the number of virtual network requests arriving is increasing, and the cost of physical network is also increasing, but the overall trend is relative steady. Among them, the STEC-IoT algorithm proposed by us has the lowest mapping cost compared with the other three algorithms. This is because the algorithm proposed in this paper considers not only the security attributes and other attributes of nodes, but also the distance between nodes and boundary nodes in the node mapping stage, which makes the cost of nodes and links at a low level, thus effectively reducing the cost.

\textbf{Experiment 4:} The comparison of substrate network revenue/cost ratio.

\begin{figure}[!htp] 
\includegraphics[width=1.0\columnwidth]{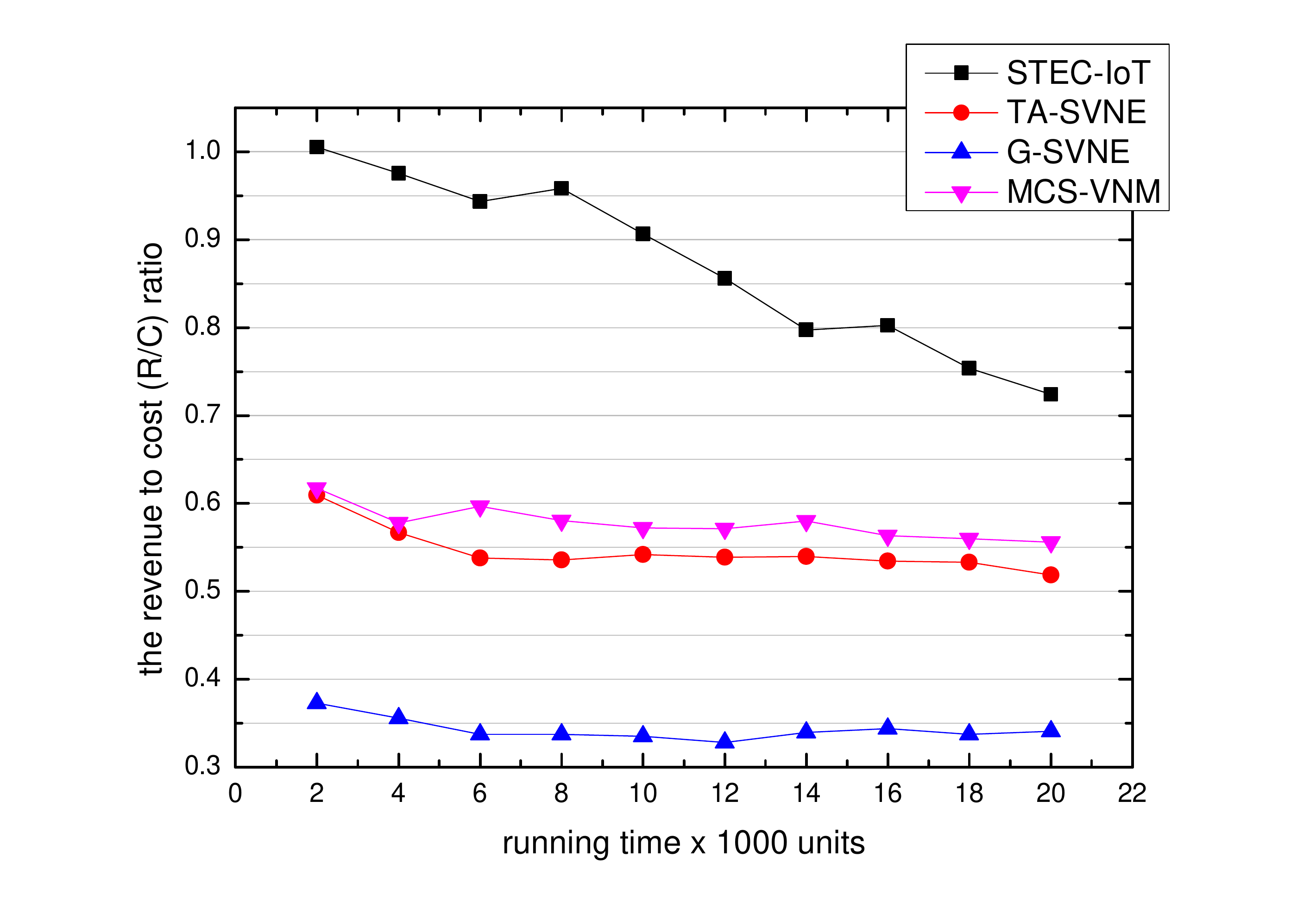}
\caption{The revenue/cost ratio of substrate network.}
\label{ex4}
\end{figure}

As shown in Fig.\ref{ex4}, we compare the revenue/cost ratio of the three algorithms. The data is the ratio of revenue and cost, which can well reflect the performance of the algorithm. We can make an observation in the Fig.\ref{ex4} that the proposed STEC-IoT algorithm in this paper has a higher ratio, which has been maintained at around 0.8 steadily, shows that the mapping revenue of the algorithm can make up for the mapping cost to a large extent, while the ratio of TA-SVNE and MCS-VNE algorithm is between 0.5 and 0.6, indicating that the mapping revenue accounts for about half of the cost, and the ratio of G-SVNE is between 0.3 and 0.4, indicating that the mapping revenue of the algorithm is difficult to maintain the cost of the mapping process.

Compared with other algorithms, STEC-IoT algorithm considers all attributes of nodes comprehensively in the node mapping stage, achieves load balancing, and uses PSO algorithm in the link mapping stage to achieve global optimization. Therefore, compared with the other three algorithms, the algorithm proposed in this paper has obvious advantages in all aspects.

\section{Conclusion}
The rapid growth of Internet of things equipment brings a significant burden to the network, and the traditional centralized public cloud service mode also faces great security risks. The edge computing based on Internet of things can efficiently process the enormous item generated by Internet of things facilities. For the complex data and the security risks faced in the process of data transmission, we propose an architecture to solve the security of edge computing by virtualizing the edge nodes. First, we propose an edge node division strategy to abstract the edge network as a virtual network. Second, we propose a security mechanism to measure the security level of nodes, finally, the data transmission process is modelled as a virtual network mapping model. The experiments in section 6 proved the advanced nature of the architecture.

In future research, we will further consider the combination of edge computing and virtual network, and use network virtualization technology to solve the problems encountered by edge computing and the Internet of things.
\bibliography{references}
\end{document}